%% file: chi24_lbw.tex
\documentclass[sigconf]{acmart} 

\AtBeginDocument{%
  \providecommand\BibTeX{{%
    \normalfont B\kern-0.5em{\scshape i\kern-0.25em b}\kern-0.8em\TeX}}}

\copyrightyear{2024} 
\acmYear{2024} 
\setcopyright{rightsretained} 
\acmConference[CHI EA '24]{Extended Abstracts of the CHI Conference on Human Factors in Computing Systems}{May 11--16, 2024}{Honolulu, HI, USA}
\acmBooktitle{Extended Abstracts of the CHI Conference on Human Factors in Computing Systems (CHI EA '24), May 11--16, 2024, Honolulu, HI, USA}
\acmDOI{10.1145/3613905.3651042}
\acmISBN{979-8-4007-0331-7/24/05}

\newcommand{\sysname}{SLA}
\newcommand{\chatsysname}{OHA}

\newcommand{\bosch}{
    \affiliation{
      \institution{Bosch Research North America}
      \city{Sunnyvale}
      \state{California}
      \country{USA}
      \postcode{94089}
    }
}
\usepackage{tabularx} % For tables that span the \textwidth and have defined column widths
\usepackage{array}    % For custom column types
\newcolumntype{M}{>{\centering\arraybackslash\bfseries}m{0.25\textwidth}}
\newcolumntype{N}{>{\raggedright\arraybackslash}m{0.375\textwidth}}
\usepackage{booktabs} % Allows for the top and bottom rule
\usepackage{ragged2e} % Allows for justified text
\newcolumntype{Y}{>{\RaggedRight\arraybackslash}X} % Defines a new column type

\usepackage{xcolor}
\definecolor{blue}{RGB}{44, 102, 194}
\usepackage{multirow} % multirow tale
\usepackage{graphicx} % subfigure
\usepackage{caption} % subfigure
\usepackage{subfig} % subfigure
\usepackage{listings}

\lstdefinelanguage{json}{
    basicstyle=\normalfont\ttfamily,
    numbers=left,
    numberstyle=\scriptsize,
    stepnumber=1,
    numbersep=8pt,
    showstringspaces=false,
    breaklines=true,
    frame=lines,
    backgroundcolor=\color{white},
    stringstyle=\color{black},
    keywordstyle=\color{blue},
    commentstyle=\color{olive},
    string=[s]{"}{"},
    comment=[l]{//},
    comment=[s]{/*}{*/},
    morecomment=[s]{\#}{\#},
    keywords={true, false, null},
    emph={},
    emphstyle=\color{red}
}
\lstdefinestyle{python}{
    language=Python,
    basicstyle=\ttfamily\small,
    commentstyle=\color{olive},
    keywordstyle=\color{blue},
    numberstyle=\tiny\color{gray},
    numbers=left,
    otherkeywords={self},
    showspaces=false,
    showstringspaces=false,
    showtabs=false,
    tabsize=2,
    frame=single
}

\begin{document}

\title[Investigating Interaction Modes and User Agency in Human-LLM Collaboration for Domain-Specific Data Analysis]{Investigating Interaction Modes and User Agency in Human-LLM Collaboration for Domain-Specific Data Analysis}


\author{Jiajing Guo}
% \orcid{0000-0003-0511-136X}
\email{jiajing.guo@us.bosch.com}
\bosch{}

\author{Vikram Mohanty}
% \orcid{}
\email{vikrammohanty@acm.org}
\bosch{}

\author{Jorge Ono Piazentin}
\email{jorge.piazentinono@us.bosch.com}
\bosch{}

\author{Hongtao Hao}
\authornote{This work was completed during an internship at Bosch Research North America.}
\email{hongtao.hao@us.bosch.com}
\bosch{}

\author{Liang Gou}
\email{liang.gou@us.bosch.com}
\bosch{}

\author{Liu Ren}
\email{liu.ren@us.bosch.com}
\bosch{}

\renewcommand{\shortauthors}{Guo and Mohanty, et al.}

\begin{abstract}
Despite demonstrating robust capabilities in performing tasks related to general-domain data-operation tasks, Large Language Models (LLMs) may exhibit shortcomings when applied to domain-specific tasks.
We consider the design of domain-specific AI-powered data analysis tools from two dimensions: 
\textbf{interaction} and \textbf{user agency}.
% The interaction with an AI-powered analysis tool can be open-ended or structured, depending on whether users can decide what they want to say or if there are predefined steps. 
% A user can have high or low agency on the output of LLM on the execution level or on the planning level.
% Leveraging the few-shot prompting technique, we introduce \agentname{}, an LLM-powered agent specifically tailored for domain-specific data analysis, which can:
% (1) augment and enrich domain knowledge provided by domain experts,
% (2) utilize relevant exemplars as prompts,
% (3) iteratively execute Python code to autonomously generate insightful analyses.
% In a preliminary sanity check study, we found that LLMs have the ability to acquire domain knowledge from exemplars.
We implemented two design probes that fall on the two ends of the two dimensions:
an \underline{o}pen-ended \underline{h}igh \underline{a}gency (\chatsysname{}) prototype and 
a \underline{s}tructured \underline{l}ow \underline{a}gency (\sysname{}) prototype.
We conducted an interview study with nine data scientists to investigate (1) how users perceived the LLM outputs for data analysis assistance, and (2) how the two design probes, \chatsysname{} and \sysname{}, affected user behavior, performance, and perceptions.
Our study revealed insights regarding participants' interactions with LLMs, how they perceived the results, and their desire for explainability concerning LLM outputs, along with a noted need for collaboration with other users, and how they envisioned the utility of LLMs in their workflow.

% Our study revealed insights regarding participants' interactions with LLMs, how they perceived the results, 
% and their desire for explainability concerning LLM outputs, along with a noted need for collaboration with other users, and how they envisioned the utility of LLMs in their workflow. Building upon these insights, we propose enhancements for \sysname{} and design implications for building an \emph{expert--LLM--novice} collaborative platform for supporting domain-specific data analysis tasks.

% To further investigate how to design a 
% human-AI collaborative system that facilitates knowledge sharing between experts and novices via \agentname{},
% we developed a web application of \agentname{} as an exploratory design probe. 
% Through qualitative interviews with eight data scientists, we identified design implications for constructing an LLM-enhanced collaborative system.

% As a sanity check study, we found LLMs can "learn" domain knowledge from exemplars. When asked a domain-specific question, LLMs achieved higher execution rate, higher correctness, and received higher ratings for insights quality when prompted with exemplars containing domain knowledge compared with the case when prompted with generic exemplars.
% To explore how to design a human-LLM collaboration system for domain-specific data analysis, we implemented a web application of \agentname{} as a design probe. We interviewed eight data scientists and identified design implications to design a collaborative system.

% collaborative system
\end{abstract}


\begin{CCSXML}
<ccs2012>
   <concept>
       <concept_id>10003120.10003121.10011748</concept_id>
       <concept_desc>Human-centered computing~Empirical studies in HCI</concept_desc>
       <concept_significance>500</concept_significance>
       </concept>
 </ccs2012>
\end{CCSXML}

\ccsdesc[500]{Human-centered computing~Empirical studies in HCI}

\keywords{Large language model, data analysis, domain knowledge, human-AI collaboration, user agency}


%   \includegraphics[width=\textwidth]{figs/teaser.png}
%   \caption{An example of how \agentname{} answers user's domain-specific question. Here "customer risk level" is a domain concept that has its unique definition in the user's organization. \agentname{} searches database and finds a log of a similar question regarding "customer risk level". This log, including question, code, and insights, is prompted as an exemplar to help LLM conduct reliable analysis.}

% \received{20 October 2023}

\maketitle

\section{Introduction}

Domain-specific data operations encompass a range of activities, including data transformation, processing, and analysis across specialized domains, such as medicine, manufacturing, finance, sports, and beyond.
Conducting complex data operations requires a deep understanding of the specific data structures and thorough knowledge of the unique concepts and terminologies relevant to each domain.

In recent years, pre-trained Large Language Models (LLMs) have demonstrated significant capabilities in diverse data science and machine learning tasks, 
such as visualization \cite{voigt-etal-2023-vist5}, 
junior-level data analysis \cite{Cheng2023-GPT4},
classification \cite{Hegselmann2023-TabLLM}, 
and model selection \cite{Shen2023-HuggingGPT}.
Many research communities and companies started to investigate Human-LLM collaboration as the future of programming, as studies showed that LLMs can save developers' searching efforts \cite{Vaithilingam2022-Expectation}, improve productivity and developer happiness \cite{Kalliamvakou2022-Copilot}. 
Commercially available tools such as Github \textit{Copilot} and ChatGPT \textit{Code Interpreter} \cite{Lu2023-Code} garnered significant attention upon their release.
% ChatGPT launched a new feature Code Interpreter in July 2023, which allows users to upload files and achieve tasks such as document summarization, data analysis, and image editing \cite{}. 
% Many user experiments and studies have shown great performance in data analytics. 

Despite demonstrating robust capabilities in general-domain data operations, LLMs may exhibit shortcomings when applied to domain-specific tasks. 
While the extensive training data enables LLMs to comprehend a wide range of concepts or terms, it may still be insufficient for understanding meanings specific to certain organizations or nuanced contexts.
% ========= Added for LBW
% Techniques such as parameter-efficient fine-tuning \cite{}
% and retrieval augmented retrieval \cite{Gao2023-RAG} have been proposed to mitigate the disadvantages of pretrained LLMs and have achieved significant performance improvements.
% However, in real-world human-AI interaction, it is common that user requests encounter the limits of LLMs' knowledge, making human users' "teaching" or "instructing" an inevitable aspect of enhancing model performance.

\input{01_table_interaction_agency}

We consider the design of AI-powered domain-specific data analysis from two dimensions: 
\textbf{interaction} and \textbf{user agency}, as shown in Table~\ref{tab:dimension}.
Many conversational LLM applications offer \textbf{open-ended interactions}, requiring users to consider their prompts to LLMs, with the freedom to use code, natural language commands, or questions.
\textbf{Structured interaction} typically features predefined stages and ensures that the user's journey through a system follows a logical path.
Users are guided by the system to complete certain steps to get results.
\textbf{User agency} or human control is a critical aspect in automated systems~\cite{heer2019agency, shneiderman2022human}.
Code assistants such as Github \textit{Copilot} offer suggestions, leaving it up to the user to decide which code snippet to use, depending on the user's actions like code execution to achieve the result.
In contrast, LLM agents such as ChatGPT \textit{Code Interpreter} execute the code on behalf of the user and automatically present the results.
% The emergence of LLM agents in recent years has made end-to-end data analysis more accessible to naive users. 
% ChatGPT Code Interpreter~\cite{Lu2023-Code} provides an automatic pipeline that includes code generation, execution, and result presentation.
AI-powered tools for data analysis such as PandasAI~\cite{PandasAI} and VizGPT~\cite{VizGPT} provided automatic workflow to convert user queries into visualization or analysis results.
Besides user agency on the \textbf{code execution} level, \textbf{planning-level} assistance is also a critical aspect in data analysis and has also been discussed~\cite{Gu2023-Wizard}.
However, little work has explored structured interaction in AI-powered data analysis tools. 
In the context of \textbf{domain-specific} tasks, human-LLM collaboration is essential as LLMs often lack sufficient domain knowledge. 
% Many novel design proposals are designed based on low user agency mindset~\cite{Mcnutt2023-Code, Gu2023-Wizard}
% Automated tools such as AutoDS~\cite{Wang2021-AutoDS} and AutoML~\cite{Weidele2020-AutoAIViz} provide a structured service that a user is guided to upload data, select an ML task configuration, and monitor the result. The system optimizes the sequence of the model pipeline, which includes data preprocessing, feature engineering, algorithm selection, and hyperparameter optimization.

% Both researchers and practitioners have explored the design space of AI-powered code assistance for data analysis~\cite{Gu2023-Wizard, Mcnutt2023-Code}.
% ~\citet{Mcnutt2023-Code} surveyed 14 interaction designs in notebooks and probed data analysts with design prototypes with a variety of features, such as \textit{Targeted cell}, \textit{Inline suggestions}, and \textit{Side panel}.
% ~\citet{Gu2023-Wizard} investigated analysts' preferences on planning assistance via a Wizard-of-oz study.

% Although free-form interaction such as Github Copilot provides such freedom to analysts, it requires the users to be equipped with programming skills.
% Domain-specific analysis often involves both data scientists and domain experts, whose programming skills vary.

% \begin{figure}
%     \centering
%     \includegraphics[width=0.8\textwidth]{figs/structure_agency_spectrum_20240125_01.png}
%     \caption{The two dimensions of AI-powered data analysis: workflow structure (open-ended v.s. structured) and user agency (low v.s. high).
%     }
%     \label{fig:system_ui}
% \end{figure}

In this work, we explore the design space of domain-specific data analysis by inviting users to work with two probes from the ends of the two dimensions: 
one has an open-ended interaction workflow and gives users high agency on both execution and planning level; 
the other has a structured interaction workflow that guides the user step-by-step and gives lower execution and planning agency compared with the first one.
We conducted an interview study with nine participants and invited them to complete two domain-specific analysis tasks on each prototype.

We aim to answer the following research questions: 

\begin{itemize}
    \item \textbf{RQ1}: How do users perceive the outputs of LLMs in the context of domain-specific data analysis tasks?
    \item \textbf{RQ2}: What is the optimal design for LLM-powered tools to assist in data analysis tasks?
\end{itemize}

Our key findings are:
\begin{itemize}
    \item Participants expressed overall satisfaction with the LLMs in assisting them with data analysis tasks, while raising concerns about the need for transparency and ability to verify results. 
    \item The different interaction modes (open-ended and structured workflow) and the two degrees of agency (high and low) resulted in different user behaviors, each with its own set of advantages and limitations. 
\end{itemize}

Based on these findings, we propose design implications for designing an LLM-enhanced collaborative system for supporting data analysis tasks.

\section{Interview Study}
\label{sec:interview}

% it does not offer an interactive user interface for facilitating human participation in the data analysis process. 
% This omission is significant because it potentially sidelines the invaluable knowledge and experiential wisdom that human collaborators could contribute to the system's analytical procedure. 
% Recognizing this gap, our forthcoming line of research aims to explore the opportunities and hurdles associated with human-LLM collaboration, seeking to design an interactive system that leverages the strengths of both human expertise and LLM capabilities.

% In this study, we developed two LLM-based tools as design probes to explore how to design a future Human-LLM collaborative system. 

% \begin{itemize}
%     \item \textbf{RQ1a}: Does \agentname{} help domain experts create exemplars efficiently?
%     \item \textbf{RQ1b}: What are domain experts’ perceptions regarding the effort involved in this approach as compared to conventional prompting?
% \end{itemize}

\subsection{Design Probes}

We implemented two interactive prototypes instead of using slide-based prototypes~\cite{Mcnutt2023-Code} or Wizard-of-oz method~\cite{Gu2023-Wizard} because we want the users to experience the application as close to their daily job as possible.
We opted not to employ existing tools such as Github \textit{Copilot} 
or ChatGPT because we want the two prototypes to use the same LLM to make the output consistent.
\textit{Code Interpreter} is a commercially available tool but its data analysis requires users to upload data files, which is not applicable for enterprises that concern data security and sensitivity.
% This decision was primarily motivated by our concerns about the sensitivity of the domain-specific data involved in our research. Given the potentially confidential or sensitive nature of the data, we aimed to minimize the risks associated with external processing or storage, thereby prioritizing data security and integrity.

\begin{figure*}[h!]
    \centering
    \includegraphics[width=\textwidth]{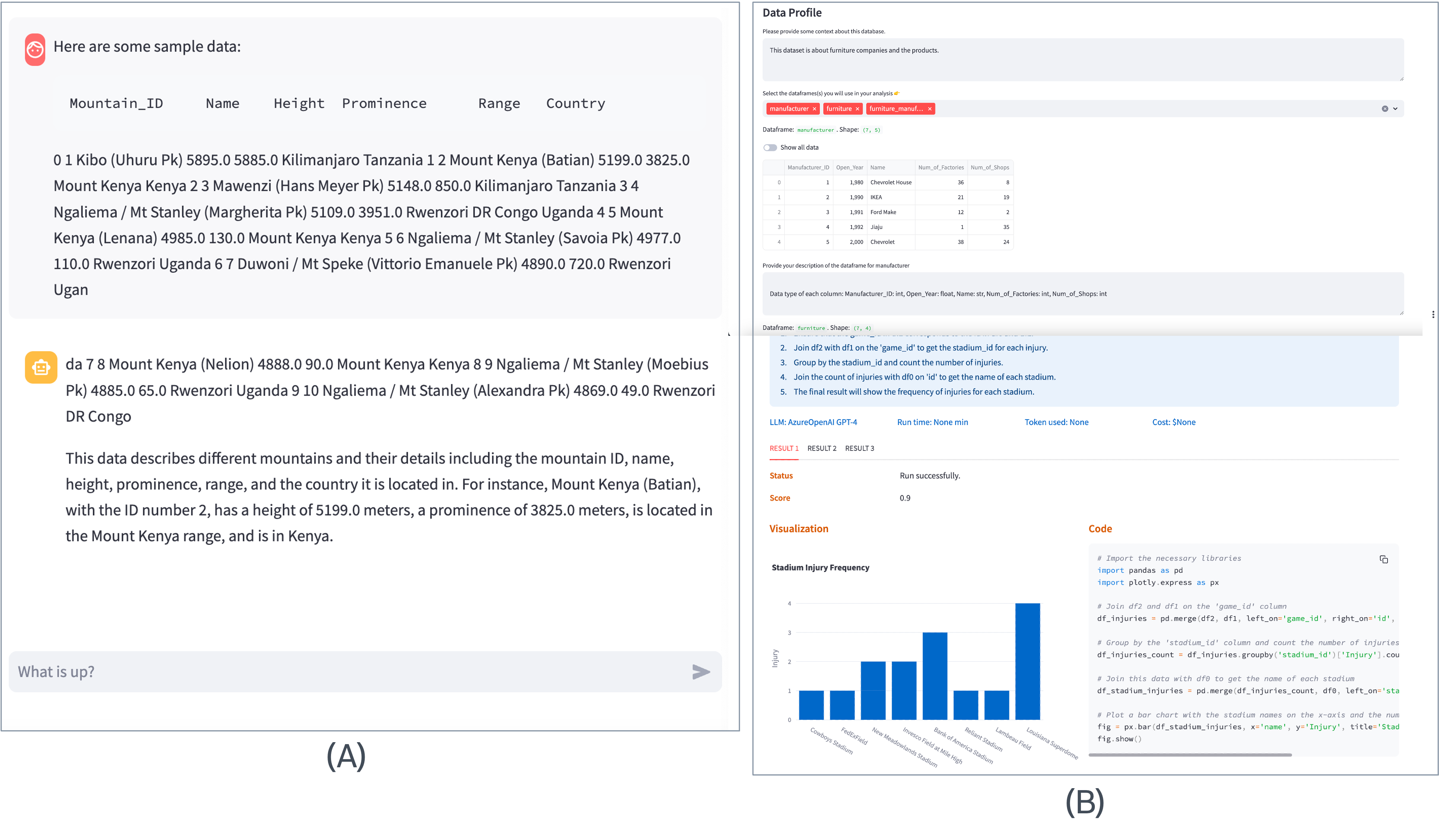}
    \caption{
    A: Interface of design probe \chatsysname{}.
    B: Interface of design probe \sysname{}.
    }
    \Description{
    A: Interface of design probe \chatsysname{}.
    B: Interface of design probe \sysname{}.
    }
    \label{fig:chat_ui}
\end{figure*}

\textbf{\chatsysname{}} (Figure~\ref{fig:chat_ui}-A) is an LLM-based chat web application integrated with a rudimentary API that connects to Azure OpenAI GPT-4 service. As a vanilla chat interface, it is limited to basic messaging functionalities. 
When used in data analysis tasks, users still need a developer environment to execute code.

\textbf{\sysname{}} (Figure~\ref{fig:chat_ui}-B) is a structured, low-agency LLM-powered application. 
It guides users through a data analysis workflow, makes plans, and executes the code on behalf of the user.
We designed the workflow based on literature about data analysis~\cite{Wongsuphasawat2019-Goals, Kandel2012-Enterprise} and LLM prompt engineering~\cite{Yao2022-ReAct, Wang2023-Plan-and-Solve}. 
The workflow is as follows:

\begin{enumerate}
    \item Describe data (User). 
    Users can upload multiple CSV files, edit descriptions of each file, and provide context information about the analysis, such as data source and domain (see Figure~\ref{fig:notebook1} in Appendix). 
    LLMs tend to perform better with more contextual information. ~\citet{Dibia2023-LIDA} examined that providing enriched data summary leads to reduced error rate when generating visualizations.
    \item Inject query and domain knowledge (User).
    Users can inject a question they want to investigate and domain knowledge that may be helpful for the LLM to complete the task.
    \item Enhance prompt (System).
    The system generates a step-by-step guide to complete the task based on the question and domain knowledge the user provided (see Figure~\ref{fig:notebook2} in Appendix).
    \item Select enhanced prompt (User).  Users can select from three variations, then the user can trigger execution.
    \item Automatic data analysis (System). This step involves code generation, execution, debugging, and insight generation. The backend is an LLM agent that adapts the ReAct framework~\cite{Yao2022-ReAct,guo2024can}, which orients the LLM to iterate through reasoning, action, and observation cycles.
    \item Review output (User). 
    Users can view the result of execution, including visualizations, code, and insights generated by \sysname{} (see Figure~\ref{fig:notebook3} in Appendix). 
    % They can rate the results from domain knowledge correctness, code correctness, and insight usefulness perspectives.
\end{enumerate}

% For example,~\citet{Dibia2023-LIDA} found that 

% We make the "enhance domain knowledge" step required as we want to collect users' opinions about this feature in a Human-LLM collaboration application. 

% \begin{figure}[h!]
%     \centering
%     \includegraphics[width=0.8\textwidth]{figs/notebook_like_3.png}
%     \caption{Interface of design probe \sysname{}.}
%     \label{fig:system_ui}
% \end{figure}

% \begin{figure*}[t!]
%     \centering
%     \begin{subfigure}[t]{0.5\textwidth}
%         \centering
%         \includegraphics[height=1.2in]{a}
%         \caption{Lorem ipsum}
%     \end{subfigure}%
    
    % \begin{subfigure}[b]{0.5\textwidth} % Specifies the width of the subfigure
    %     \includegraphics[width=\textwidth]{figs/chatgpt_like_1.png}
    %     \caption{Caption for the first figure}
    %     \label{fig:1}
    % \end{subfigure}
    % \hfill % Optional: Adds horizontal space between the figures
    % \begin{subfigure}[b]{0.5\textwidth} % Adjust the width as needed
    %     \includegraphics[width=\textwidth]{figs/notebook_like_3.png}
    %     \caption{Caption for the second figure}
    %     \label{fig:2}
    % \end{subfigure}
    % Optional: Caption and label for the entire figure environment
%     \caption{Caption for both figures}
%     \label{fig:both-figures}
% \end{figure*}

\subsection{Tasks and Procedure}

Following a lab experiment study guideline, we conducted a within-subject study to compare the interactive system built upon our framework and existing LLM applications.

% \textbf{Tool 2: \sysname{}}. We have introduced the \sysname{} in the section above. 

\textbf{Participants.}
We recruited nine data scientists from a product team in a multi-national enterprise (8M, 1F), aged from 25 to 46).
Their job content includes data analysis in manufacturing and sensor domains. 
All of them have at least one year of working experience in data analysis.
Participants self-identified job titles are 
data scientist (6), 
data engineer (1), 
and research scientist (2).
Participants have various degrees of familiarity with LLM applications for data analysis.
Some participants have experience of using LLMs to polish writing and summarize articles but have not used LLMs in data science job.
Some express untrust towards LLMs, and use LLMs to generate code templates or overall structures only. Two use LLMs frequently in their daily work.
% \todo{Four} participants have seen demos or read about AI-based code assistant for data analysis. 
% \todo{Two} have tried out some tools and \todo{one} use such tools at work.
% Details are in Table~\ref{}.
Each participant received \$20 gift card as compensation.

Before the study, participants completed a pre-study survey asking for demographic information and their previous experience with LLM-powered applications.
% and their expectations of such tools \cite{Ross2023-Programmer}.
The interview study can be broken down into three parts. 

\textbf{\textit{Part 1: Pre-study interview}}. 
We first asked some questions about participants' job content that is relevant to data analysis. Participants were cued to recall an experience when they needed to deliver data analysis reports to stakeholders, the volume of data to be analyzed, the negotiation procedure, and the format of the report. 

\begin{figure}[h!]
  % \begin{minipage}{0.5\textwidth}
  %   \includegraphics[width=\linewidth]{figs/interview_procedure.png}
  %   \caption{Interview procedure}
  % \end{minipage}%
  \begin{minipage}{0.5\textwidth}
    \includegraphics[width=\linewidth]{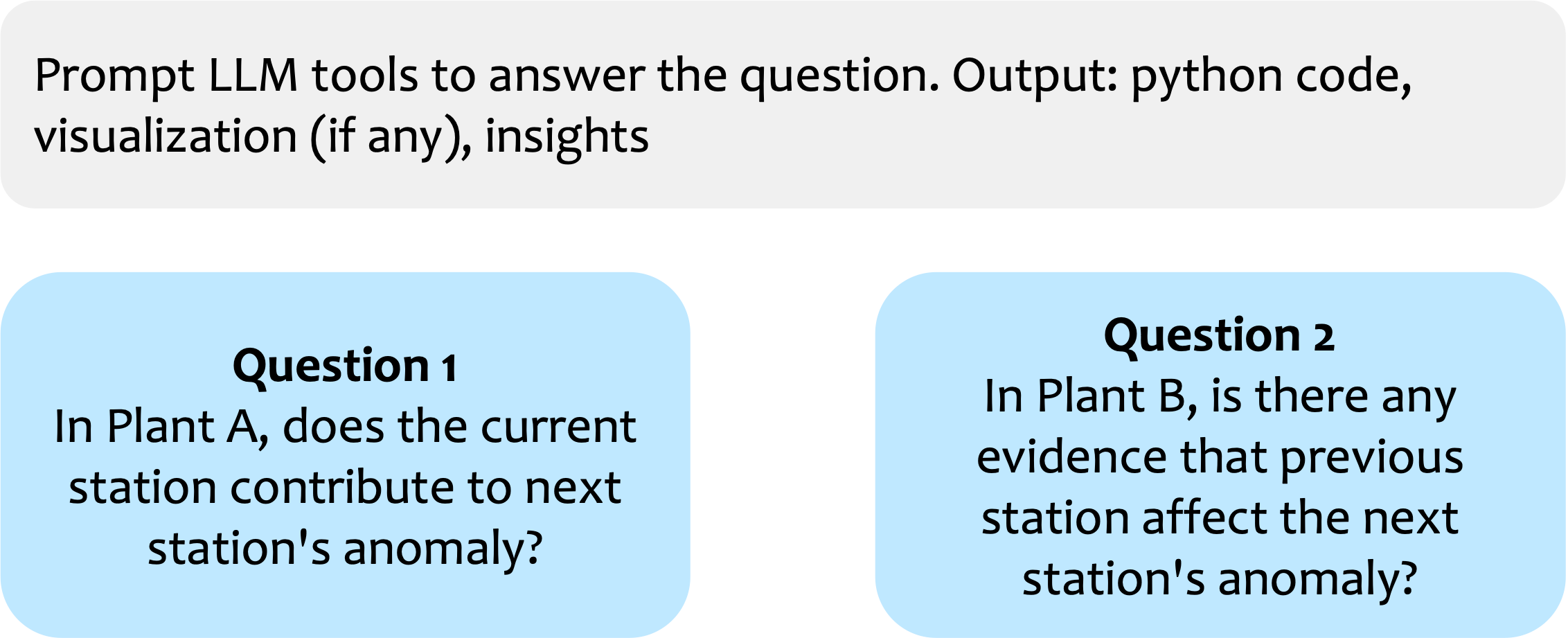}
    \caption{Example tasks in the interview study.
    Participants received the example task as a prompt and were informed that they could customize their own queries that fit their job.}
    \Description{Example tasks in the interview study.
    Participants received the example task as a prompt and were informed that they could customize their own queries that fit their job.}
    \label{fig:interview_tasks}
  \end{minipage}
\end{figure}

\textbf{\textit{Part 2: Two domain-specific data analysis tasks}}. 
Participants were asked to complete two data analysis tasks on enterprise internal data with the help of LLM-powered tools, one is \chatsysname{}, and the other is \sysname{}. Participants were encouraged to "think aloud" while they were doing the task. 
The order of tools was randomized. 
We want to simulate participants' daily work as much as possible, therefore the data used in the study were extracted from a huge data lake to which the authors have access.
% Furthermore, the tasks were tailored to participants' job content. For example, some participants focus on outlier detection, some focus on prediction or data transformation. 
The two tasks used the same data source but different requirements (see Figure~\ref{fig:interview_tasks}).

% After each task, participants filled out a survey about their workload (NASA TLX Short) \cite{Hart1988-TLX}, technology usefulness \cite{}, interaction \cite{}, importance of the tool's abilities \cite{Ross2023-Programmer}, and experience \cite{Ross2023-Programmer}. 
% Using the survey as a probe, we asked the participants questions about their opinions on this tool, aspects they like and dislike, features that can be improved, and how the tool may be used in their daily work. 

\textbf{\textit{Part 3: Critiques and reflections}}.
After two tasks, participants critiqued and reflected on their experience by comparing the two tools, and discussing the perceived utility and limitations.
% We further presented some prototype features that were not implemented in \sysname{} prototype, such as exemplar searching and domain knowledge retrieval. 

% \begin{figure}
%   % \begin{minipage}{0.5\textwidth}
%   %   \includegraphics[width=\linewidth]{figs/interview_procedure.png}
%   %   \caption{Interview procedure}
%   % \end{minipage}%
%   \begin{minipage}{0.5\textwidth}
%     \includegraphics[width=\linewidth]{figs/interview_tasks.png}
%     \caption{Example tasks in the interview study}
%     \label{fig:interview_tasks}
%   \end{minipage}
% \end{figure}

% \begin{figure}
%     \centering
%     \includegraphics[width=0.8\textwidth]{figs/interview_procedure.png}
%     \caption{Interview procedure}
%     \label{fig:interview_procedure}
% \end{figure}

% \textbf{Data Analysis}.
% Besides post-task surveys, we also collected the following data:
% \begin{itemize}
%     \item User behavior data coded from video recordings with participants’ consent.
%     \item The final exemplars selected by the participants;
%     \item Post-study questionnaire;
%     \item Post-study semi-structured interview;
% \end{itemize}

Interviews were conducted by two researchers. One researcher led the interview and the other took notes. The two researchers had retrospective meetings after each interview and discussed the takeaways.
All the interviews were video-recorded and with participants’ permission and transcribed. 
Using an inductive open-coding analysis approach \cite{Corbin2012-Grounded, saldana2021coding}, the authors read through the transcripts and watched recordings multiple times to conduct the initial coding. 
% For instance, we identified codes such as "benefits of expanding one’s own practice" to describe a teacher’s motivation for conducting live meditation, "energy in in-person sessions" and "convenience of live sessions" to highlight the differences between live and in-person sessions, "requesting donations" and "distractions from the message during the session" to describe the challenges that teachers reported. 
Then, both authors read through and discussed the coded transcripts together, iteratively highlighted excerpts, and identified key insights and recurring patterns in the data. 
% We use pseudonyms (see Table 1) to label quotes from participants’ transcripts.

% We followed the guidelines of grounded theory  [19] but adapted the process to our research questions. 
% We started with an initial open coding and used memoing techniques to inductively develop an inventory of codes. Then the authors discussed the labels and related data by categorizing and merging similar ones. For example, we grouped all  the codes about instructors’ challenges. We used axial coding to construct the linkages between data. 
% Several iterations were conducted to identify emerging themes such as the challenge of real-time interaction  and exercise. We stopped  the  data  collection as  we  reached  saturation  and  no  new  themes emerged. In the end, selective coding was employed to generate concepts and themes and  present the results.

\section{Preliminary Findings}
\label{sec:findings}

In this section, we discuss the insights from our probe interviews, which we have organized according to our two main research questions: (1) how users perceived the LLM outputs for data analysis assistance, and (2) how the two design probes, \chatsysname{} and \sysname{}, affected user behavior, performance, and perceptions. 

\input{chi24_lbw_findings}

\section{Discussion}

The design probes used in the interview were designed and implemented to align with the two ends on the \textbf{interaction} and \textbf{user agency} dimensions.
In this section, we discuss the findings and design considerations from these dimensions. 
Then we discuss the trust and verifiability of LLM-powered data analysis tools and design recommendations.

\subsection{Design Considerations of AI Assistance for Domain-specific Data Analysis}

% We identified two core areas where AI assistants can enhance data analysis. 
% First, there's a need for general-purpose support, covering tasks from basic data processing to brainstorming analysis questions and assisting in presentation. 
% Second, a set of domain-specific design considerations emerged, highlighting the importance of leveraging past domain knowledge and analyses, understanding dataset intricacies, and receiving on-demand tailored guidance throughout the analysis. 
\textbf{Structured Interaction}.
In HCI literature, structured workflows have been widely adopted in interactive applications. This approach, which predates the advent of LLM applications such as ChatGPT, is designed to streamline user workflows and guide users to prevent errors. Since the launch of ChatGPT, conversational and open-ended interaction have grown in popularity, offering more flexible and natural interaction paradigms. Many AI-powered tools for data analysis~\cite{PandasAI, VizGPT} and research-oriented design prototypes have adopted the open-ended interaction paradigm~\cite{Mcnutt2023-Code, Gu2023-Wizard}.
Despite allowing users the freedom to initiate conversations, open-ended interaction necessitates that users possess prompt engineering skills, such as choice of terminology, arrangement of word order, as well as repetition in keywords~\cite{Zhou2023-Scaffolding}.
This requirement may burden data analysts with the additional task of mastering prompt engineering if their original goal of using AI-powered tools is to increase efficiency~\cite{Mcnutt2023-Code, Barke2023-Grounded}.
For users inexperienced with prompt engineering, predefined steps aid in understanding how LLMs function. Intermediate steps elucidate how the LLM interprets tasks and decides on actions, enhancing transparency.
The appropriateness of conversational interaction for LLM-powered data analysis remains an open question. Designers and researchers should avoid creating chatbots merely for their own sake.
Instead, considering a system that incorporates both open-ended and structured interactions may be beneficial.
For instance, developers could implement a conversational assistant that guides users through predefined steps, or a structured interface that allows for configurable steps.

% usually has predefined stages and ensures that the user's journey through a system follows a logical path.
% Users are guided by the system to complete certain steps to get results.
% Structured interaction with pre-defined stages is common in non-AI systems. 

% ~\cite{lee2020solutionchat}.

% Zhou2023-Scaffolding
% (1) choice of terminology, arrangement of word order, as well as repetition in keywords can infuence whether a prompt can generate functionally correct code;
% (2) insufcient interaction between Human and GitHub Copilot could make users over-trust or under-trust the ability of GitHub Copilot. This results in the failure of data analysis tasks;
% (3) GitHub Copilot has its faws such as ignoring information in the prompts given by users. Future improvements can be targeted on more precise information capture.

% understanding, editing, and debugging code snippets generated~\cite{Vaithilingam2022-Expectation}

% Low agency in execution is fine.
\textbf{Execution-level User Agency}.
Participants expressed appreciation for the automated processes in code generation, debugging, and interpreting results.
This finding suggests that low levels of control at the execution level do not result in dissatisfaction among our participants.
It was suggested that increased agency, allowing data analysts to modify the code before and after execution, would be beneficial.
We also found that depending on the job content, the range of execution-level tasks may be expanded to data wrangling pipeline instead of only code generation and execution.
Participants expressed \textbf{a need for automation in foundational data pre-processing and wrangling tasks}, such as normalizing data, filling in missing values, extracting descriptive statistics, finding outliers, and visualizing distribution.
These tasks, despite their repetitive nature, are crucial for thorough analysis~\cite{felix2019systematic}.
Our observations confirmed that participants consistently prioritize these initial steps, underscoring the potential benefits of streamlining them through automation~\cite{anaconda2020}.
The efficacy of LLMs in automating fundamental pre-processing tasks streamlines initial data analysis stages~\cite{kandel2011wrangler, karmaker2021automl, bilal2022auto, heer2019agency}, enabling human analysts to focus on complex, domain-specific analytical challenges.
% In fact, they represent set processes that analysts invariably undertake before delving into more intricate analyses~\cite{endel2015data, fan2021review}. 
% This segues into the necessity of \textbf{LLMs accessing data for contextual understanding}, 
% and \textbf{our proposal for a data wrangler module with LLM support}, inspired by AutoDS~\cite{wang2021autods}, to adeptly perform these operations. 

% planning level agency is needed.
\textbf{Planning-level User Agency}.
In contrast to their significant enthusiasm for assistance at the execution level, participants showed interest and caution regarding planning-level assistance. 
One reason for their caution is a lack of confidence in the quality of plans generated by LLMs.
While using \chatsysname{}, most participants opted to create execution commands rather than consult the LLM's analysis plan.
The planning mechanism in \sysname{}, which generates a step-by-step guide based on user questions and domain knowledge, demonstrated limited capabilities and may not suit complex real-world tasks.
Some participants were interested in whether the LLM-powered tool could assist in brainstorming potential analytical approaches and facilitating iterative exploration. 
However, they emphasized that the tool should offer suggestions, while the final decision on which suggestion to adopt should rest with them as data scientists.
This finding underscores the importance users place on the planning-level agency. It is imperative that design considerations reflect this priority.

\subsection{Calibrating User Trust and Expectations with AI Explainability in Data Analysis Assistance}

During our probe, \textbf{participants exhibited a general caution towards the results} generated by \sysname{}, mainly due to the \textbf{the absence of methods to verify these outcomes} and an expectation for \textbf{consistent results}. 
Additionally, they also raised questions concerning the process by which the LLM reached specific conclusions, impacting their trust in the outputs of LLM-based systems collectively. 
While this behavior aligned with prior work on data scientists seeking interpretable results~\cite{kaur2020interpreting}, this skepticism could be rooted in the well-documented hallucination tendencies of LLMs~\cite{ji2023survey}.
It could be beneficial to offer users a verification sandbox playground, allowing them to manually conduct their analyses and compare results with those generated by the LLM, aiding in \textbf{calibrating expectation and understanding the capabilities of an LLM-based system}. 
% Moreover, the inherent collaborative framework of our platform could serve as a conduit for \textbf{users to seek peer validations}, enhancing trust in the analyses conducted~\cite{ehsan2021expanding}.

The quest for transparency was evident when participants, with enhanced domain knowledge, showed a preference for step-by-step elucidation of the analyses, assuming that \sysname{} followed these steps. 
This hints at the potential benefit of \textbf{visualizing the analytical process through a flowchart} or similar means, displaying intermediate steps, and even showcasing counterfactual results at each juncture for better user validation. 
% Participants favored \chatsysname{} interface, enjoying control over code execution, unlike other conditions where the tool autonomously handled errors, creating a disconnect and confusion~\cite{wang2019human}. 
% As underscored by prior work, \textbf{user agency is pivotal} in automated systems~\cite{heer2019agency, shneiderman2022human}; 
% an IDE integration or a DomainDA-enhanced notebook could foster this agency, facilitating on-demand assistance during analysis. 
Enhancing transparency --- visualizing backend operations and ensuring step-by-step explainability --- can demystify the process, nurturing trust and engagement with the system. 
% Incorporating data analysis assistants' reliance on domain-specific exemplars, another layer of transparency could be added by \textbf{revealing to the user the source information} -- such as the particular analysis code or domain knowledge utilized. 
% This, while also affording the user the ability to modify the source, fosters a clearer understanding and a more interactive engagement with the system.

\section{Limitations and Future Work}
In this study, we only interviewed data scientists with experience with manufacturing and transportation data. As some participants pointed out, the tool can be beneficial for users who have rich domain knowledge but little programming skills.
Future work would propose new design ideas with the suggestions we list above. We also would like to consider different design options for users with different skills and needs.
% Next step, we will iterate the \sysname{} and invite domain experts to try the tool.

\section{Conclusion}

We explored the design of AI-powered, domain-specific data analysis tools through two distinct prototypes, differing in interaction styles and user agency, and conducted interviews with data scientists to assess their effectiveness. Our findings showed that while users appreciated the assistance of LLMs, they also emphasized the need for transparency, result validation, and different interaction workflows catered to varying levels of agency. This research opens the door for future studies on the integration of LLMs into domain-specific applications, focusing on enhancing user collaboration, explainability, and adaptable interaction modes.

% \subsection{}
% Much like other LLM-based tools, result hallucination is a major issue for this tool. 
% However, the participants we interviewed were data scientists with a background in Computer Science and were largely familiar with the nature of LLMs.

% \todo{Required by ACM}

% \section{Conclusion}
% We 

% \begin{acks}
% \end{acks}


\bibliographystyle{ACM-Reference-Format}
\bibliography{main}

\clearpage
\appendix

% \begin{figure}[h!]
%   % \begin{minipage}{0.5\textwidth}
%   %   \includegraphics[width=\linewidth]{figs/interview_procedure.png}
%   %   \caption{Interview procedure}
%   % \end{minipage}%
%   \begin{minipage}{0.5\textwidth}
%     \includegraphics[width=\linewidth]{figs/interview_tasks.png}
%     \caption{Example tasks in the interview study.
%     Participants received the example task as a prompt and were informed that they could customize their own queries that fit their job.}
%     \label{fig:interview_tasks}
%   \end{minipage}
% \end{figure}

% \section{Sample Prompts}
% \label{appendix:sample_prompts}

% \begin{figure}
%     \centering
%     \includegraphics{}
%     \caption{Probe~\sysname{}}
%     \label{fig:enlarged}
% \end{figure}

\begin{figure*}[h]
    \centering
    \includegraphics[width=\textwidth]{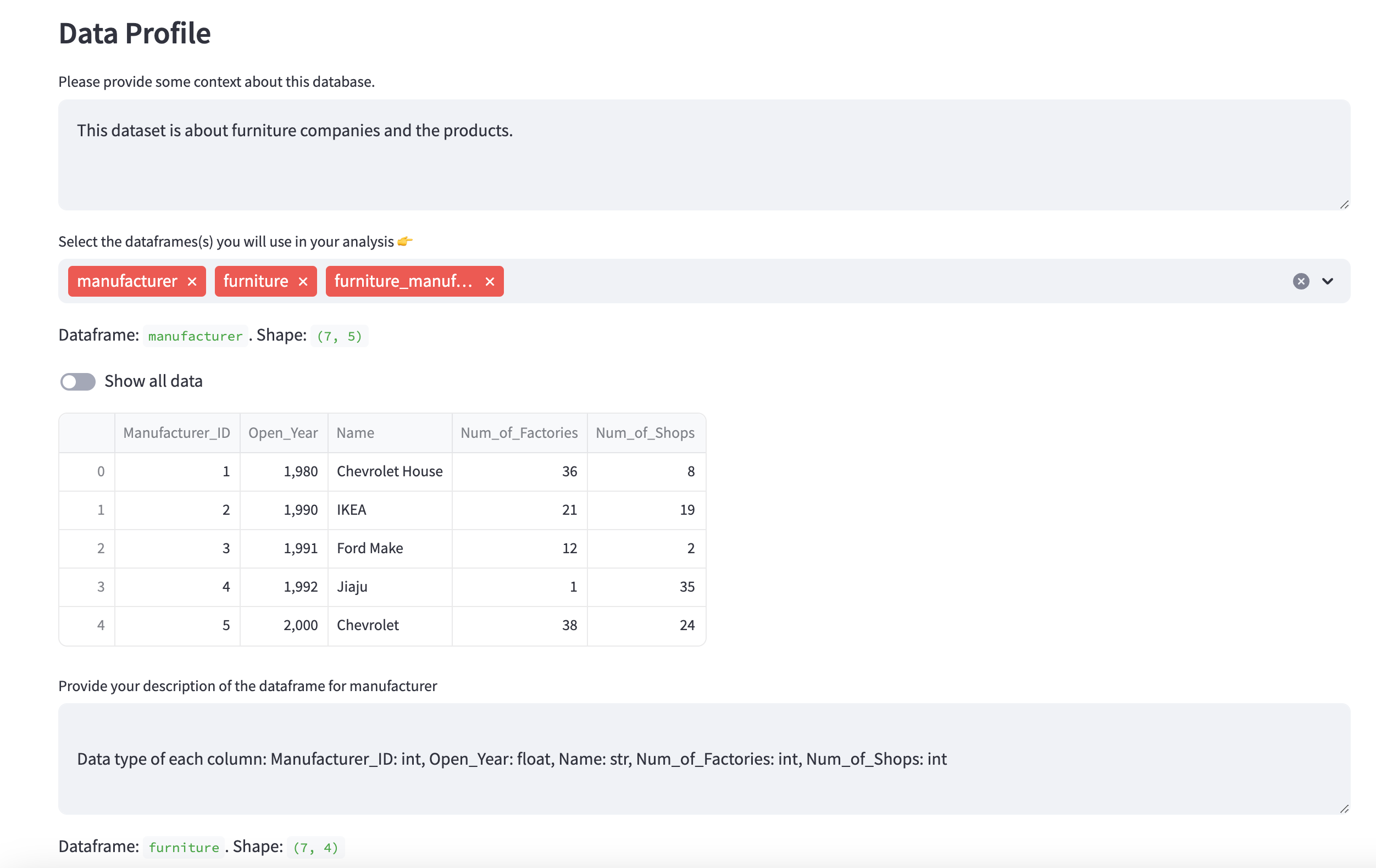}
    \caption{Probe~\sysname{}: Adding Data Context}
    \Description{Probe~\sysname{}: Adding Data Context}
    \label{fig:notebook1}
\end{figure*}

\begin{figure*}[h]
    \centering
    \includegraphics[width=\textwidth]{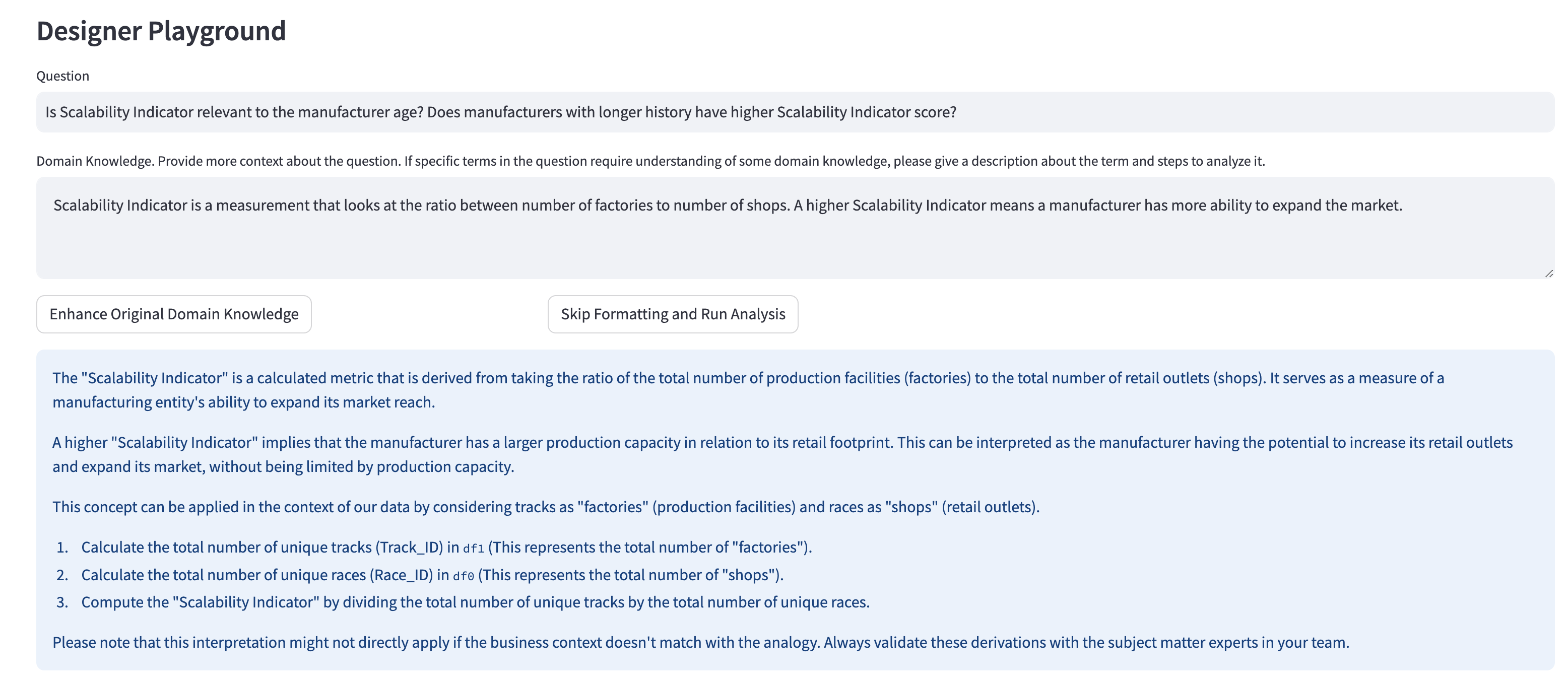}
    \caption{Probe~\sysname{}: Providing Domain Knowledge and Enhancing it with the LLM}
    \Description{Probe~\sysname{}: Providing Domain Knowledge and Enhancing it with the LLM}
    \label{fig:notebook2}
\end{figure*}

\begin{figure*}[h]
    \centering
    \includegraphics[width=\textwidth]{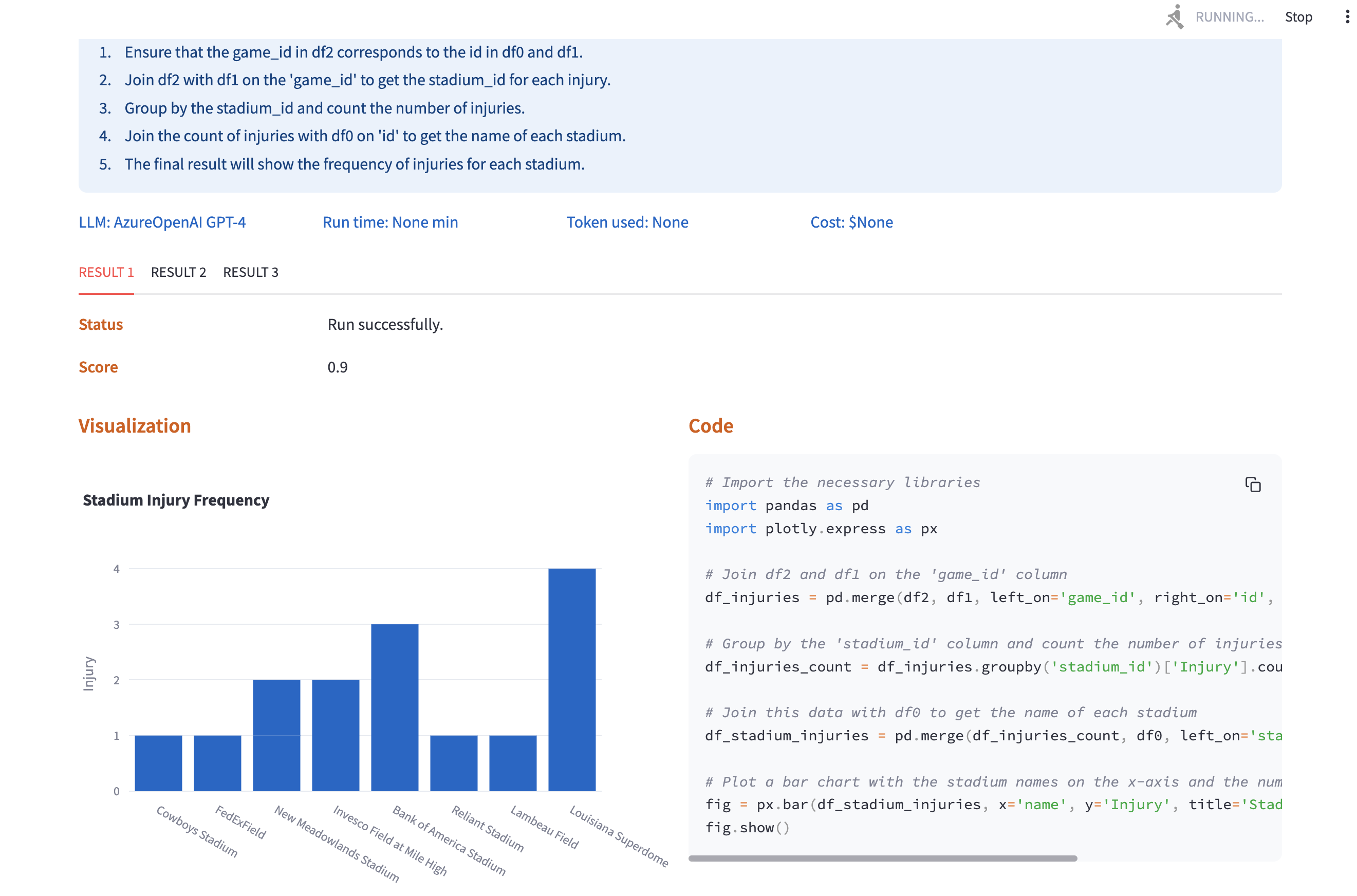}
    \caption{Probe~\sysname{}: Analysis results -- visualization and code. Insights are not shown in this figure.}
    \Description{Probe~\sysname{}: Analysis results -- visualization and code. Insights are not shown in this figure.}
    \label{fig:notebook3}
\end{figure*}

% \section{Design Probe}
% \label{appendix:design_probe}

\end{document}

%% file: 01_table_interaction_agency.tex
\begin{table*}
% [h!]
\centering
\begin{tabular}{cccc}
% {
% >{\raggedright\arraybackslash}p{0.2\textwidth}
% p{0.2\textwidth}
% p{0.2\textwidth}
% p{0.4\textwidth}
% }
\toprule
\multirow{2}{*}{\textbf{Interaction}} 
& \multicolumn{2}{c}{\textbf{User agency}} 
& \multirow{2}{*}{\textbf{Application}}
\\
% \cline{2-3}
& \textbf{Execution} & \textbf{Planning} &
\\ 
\midrule
Open-ended
& high
& high
& Copilot, ChatGPT, \textbf{*Design Probe~\chatsysname{}}
\\ 
Open-ended
& low
& high
& Code Interpreter, PandasAI, VizGPT 
\\ 
Structured
& low
& low
& \textbf{*Design Probe~\sysname{}}                     
\\ 
% Structured
% & medium
% & medium
% & Design recommendation
% \\ 
\bottomrule
\end{tabular}
\caption{
    AI-powered coding tools on interaction and user agency dimension.
    We implemented two design probes on the ends of the two dimensions: 
    an \underline{o}pen-ended \underline{h}igh \underline{a}gency (\chatsysname{}) that has an open-ended workflow and gives users high execution and planning agency on LLM outputs;
    another is a \underline{s}tructured \underline{l}ow \underline{a}gency (\sysname{}) application that guides users step-by-step and automatically plans and executes code to complete tasks.
}
\label{tab:dimension}
\end{table*}

% \begin{table*}[h] % 'h' for 'here' placement
% \begin{tabularx}{\textwidth}{|>{\centering\arraybackslash}X|c|c|>{\centering\arraybackslash}X|}
% \toprule
% \multirow{2}{*}{\textbf{Interaction Mode}} & \multicolumn{2}{c|}{\textbf{User Agency}} & \multirow{2}{*}{\textbf{Application}} \\ \cline{2-3}
%                                       & \textbf{Execution} & \textbf{Planning}    &                                      \\ 
%                                       \midrule
% Open-ended                            & High               & High                 & Copilot, ChatGPT, \textbf{*Design Probe OHA}    \\ 
% % \hline
% Open-ended                            & Low                & High                 & Code Interpreter, PandasAI, VizGPT    \\ 
% % \hline
% Structured                            & Low                & Low                  & \textbf{*Design Probe SLA}                      \\ 
% % \hline
% \bottomrule
% \end{tabularx}
% \caption{AI-powered coding tools on interaction and user agency dimension.
% We implemented two design probes on the ends of the two dimensions: 
% an \underline{o}pen-ended \underline{h}igh \underline{a}gency (ChatGPT) that has an open-ended workflow and gives users high execution and planning agency on LLM outputs;
% another is a \underline{s}tructured \underline{l}ow \underline{a}gency (SLA) application that guides users step-by-step and automatically plans and executes code to complete tasks.}
% \label{tab:dimension}
% \end{table*}

%% file: chi24_lbw_findings.tex
\subsection{User Perception of LLM outputs for Domain-specific Data Analysis}

In terms of how users perceived LLM outputs for their data analysis tasks across the two design probes, we identified three broad themes:

\subsubsection{Output Quality} 

Participants appreciated \chatsysname{} for its ability to assist in quickly generating initial versions of code, which was seen as a valuable feature for starting data analysis tasks, even though executing the code threw errors at times. 
P5 said, \textit{"It's quite easy and intuitive, especially for easy tasks, they are just supposed to write out some code."}
They also expressed similar satisfaction with the insights produced by \sysname{}, acknowledging its utility in enhancing workflow efficiency, especially during the exploratory phase and reporting stage. The prototype's ability to generate comprehensive plots and explanations was particularly valued for its potential to aid communication with stakeholders. 

\begin{quote}
    \textit{
    I think for the first time even for us to think what to tell to customer from this graph, it takes a little bit time to think how to explain. But here, if we read these insights, it gives some tips what can make sense... To start with, it is really good. Maybe a little bit more articulation or additional information might be required when we are telling to the user. I think this insights are helpful. (P8)
    }
\end{quote}

Participants expressed overall satisfaction with the code and insights produced by LLMs across both probes, even though some conclusions lacked in-depth findings or did not align with the problem. Despite these limitations, participants were optimistic about how the LLM outputs here can be a promising starting point for future iterations. 

\subsubsection{Trust and Verifiability} Trust emerged as a crucial concern among participants using \sysname{}, particularly due to challenges in verifying the tool's results. Users expressed hesitation in relying on the system without independent verification, underscoring the need for trust-building measures. P2 encapsulated a similar sentiment for \chatsysname{}: \textit{"I'm seeing how nice that was. It was really quick to get an answer. I don't know how to trust it without actually redoing all the work myself, which would be very time-consuming."} The lack of domain-specific fine-tuning in LLMs further contributed to this trust issue, with users like P1 and P4 voicing concerns about the system's understanding of their intentions and domain knowledge. Additionally, the inconsistency in the tool's outputs, attributed to the LLM's \emph{temperature} setting, was found to impact trust, especially for users less familiar with data analysis methodologies. 

\subsubsection{Need for Explanations} Participants emphasized the importance of explanations and transparency when interacting with LLMs. They actively sought clarity on the LLM's reasoning process, particularly when outputs did not match expectations. P2 captured this need: \textit{"Cause at least it gives me an idea that it understands the concept and what to do... I could edit this if I was doing a real thing and make sure that this is exactly the right steps."} This desire for explanations was especially pronounced in situations where the LLM suggested complex operations, like machine learning models, but did not follow through, leading to dissatisfaction and confusion. P7's experience underscores this frustration: \textit{"It completely over promised the delivery of the machine learning model... I'm very unsatisfied that it does not give me a model at all and also does not explain to me why."} This finding highlights a need for more transparent communication of the LLM's decision-making process to build user trust and understanding.

\subsection{Impact of Design Modes on User Behavior}

Here, we discuss how the interaction modes (i.e., open-ended and structured) and the user agency (i.e., high v/s low) in the two probes affected user behavior and perception of the LLM prototypes. 

\subsubsection{Interaction Modes} We examine the impact of the two distinct user workflows — open-ended (\chatsysname{}) and structured (\sysname{}) — on user behaviors (see Table~\ref{tab:interactionsummary}) and perceptions. In the open-ended \chatsysname{} mode, participants with LLM experience often included sample data or key columns in their queries, seeking to provide context. P4 highlighted the cumbersome nature of this process: \textit{"I wish we have a better way to directly inject this code and meta informational data into this system... I have to manually write down this prompt by myself, which is somewhat not easy."} Conversely, in the structured \sysname{} mode, users typically did not engage in an interactive dialogue about the data. Instead, they provided brief text descriptions or essential details about the dataset, relying on the system's structure to guide the data analysis process.

For domain knowledge, \chatsysname{} users often embedded this information within their questions or as side notes, while \sysname{} users provided varying levels of detail, intuitively gauging what the model might need. P1's experience illustrates this: \emph{"With the domain knowledge, I can be more clear about what features and values I want... Like traffic lights on the left, maybe the language model would interpret this as high values for left, but instead, this is low values for left."} This varied approach to input in \sysname{} sometimes led to users selecting the first provided LLM-enhanced domain knowledge variation and proceeding without exploring other options, citing time constraints. This structured approach in \sysname{} helped guide users in providing necessary information, sometimes filling in gaps that were more challenging in the open-ended \chatsysname{} mode.

\subsubsection{Agency} In the \chatsysname{} setup, users demonstrated high agency by independently determining the course of their analysis, making decisions without system guidance. In contrast, the \sysname{} probe guided users through the process, providing detailed domain knowledge and outlining analysis steps, thus reducing the level of user agency in planning. Error handling in \chatsysname{} demonstrated high user agency, with participants actively diagnosing and communicating errors to the LLM. Users copied the entire or partial error messages, directly engaging in the debugging process. In contrast, the \sysname{} probe exhibited low user agency in error handling; most errors were automatically managed within the workflow, often remaining unseen by users. 

However, in cases of unresolved errors due to token limitations, user agency was limited to rephrasing inputs without direct involvement in error resolution. Participants across both setups underscored the need for iteration capabilities in LLM-powered tools: \emph{"I think it does need some iteration with the model, and then after you take the output of the model, you need to adapt it a little bit on your own."} 

\begin{table*}[h]
\centering
\resizebox{\textwidth}{!}{%
\begin{tabular}{>{\raggedright\arraybackslash\bfseries}m{0.2\textwidth}m{0.4\textwidth}m{0.4\textwidth}}
\toprule
\textbf{Interaction} & \textbf{\chatsysname{} Probe} & \textbf{\sysname{{} Probe}} \\ \midrule
Data Context Establishment & 
  Copied sample data or columns from Jupyter notebok, e.g., ``Write python code that takes a pandas dataframe with the following fields <copied\_text>." \newline
  Drafted prompt in a notepad then paste to chat textbox. \newline
  Only described useful columns. \newline
  Directly asked questions without giving any information about the data. & 
  Selected dataset or upload own csv files. \newline
  Manually typed column description in the data description text box. \\ \midrule
Domain Knowledge Injection & 
  Manually typed in the text box. \newline
  Instead of giving a formal definition of domain concepts, directly gave concrete instructions of calculation or plotting. & 
  Manually provided input and generated variation(s). \newline
  Identified and corrected incorrect augmented domain knowledge. \\ \midrule
Error Handling and Debugging & 
  Copied the partial or entire error message from Jupyter notebook. \newline
  Tried to debug by themselves & 
  Modified the question or domain knowledge and run analysis again. \\ \midrule
Output Interpretation & 
  Read the plot or transformed data, gave explanation based on their own interpretation. & 
  Read through the "insights" generated by LLM, critiqued and revised what is wrong. \\ \bottomrule
\end{tabular}%
}
\caption{Summary of participants' interactions with the two LLM-based tools used in the interview study.}
\label{tab:interactionsummary}
\end{table*}